\documentclass[12pt,a4paper]{article}
\usepackage[utf8]{inputenc}
\usepackage[english]{babel}
\usepackage{amsmath}
\usepackage{amsfonts}
\usepackage{amssymb}
\usepackage{graphicx}
\usepackage{tikz}
\usepackage{algorithm}
\usepackage{algorithmic}
\usepackage{hyperref}
\usepackage{cite}
\usepackage{multirow}
\usepackage{booktabs}
\usepackage{geometry}
\usetikzlibrary{shapes,arrows,positioning,calc}

\title{EarthOL: A Proof-of-Human-Contribution Consensus Protocol - Addressing Fundamental Challenges in Decentralized Value Assessment with Enhanced Verification and Security Mechanisms}

\author{Jiaxiong He}
\date{\today}

\begin{document}
	
	\maketitle
	
	\begin{abstract}
		This paper introduces EarthOL, a novel consensus protocol that attempts to replace computational waste in blockchain systems with verifiable human contributions within bounded domains. While recognizing the fundamental impossibility of universal value assessment, we propose a domain-restricted approach that acknowledges cultural diversity and subjective preferences while maintaining cryptographic security. Our enhanced Proof-of-Human-Contribution (PoHC) protocol uses a multi-layered verification system with domain-specific evaluation criteria, time-dependent validation mechanisms, and comprehensive security frameworks. We present theoretical analysis demonstrating meaningful progress toward incentive-compatible human contribution verification in high-consensus domains, achieving Byzantine fault tolerance in controlled scenarios while addressing significant scalability and cultural bias challenges. Through game-theoretic analysis, probabilistic modeling, and enhanced security protocols, we identify specific conditions under which the protocol remains stable and examine failure modes with comprehensive mitigation strategies. This work contributes to understanding the boundaries of decentralized value assessment and provides a framework for future research in human-centered consensus mechanisms for specific application domains, with particular emphasis on validator and security specialist incentive systems.
		
	\end{abstract}
	
	\section{Introduction}
	
	\subsection{Technology Convergence and Current Limitations}
	
	The convergence of social computing, gamification technology, and blockchain innovation over the past two decades has created unprecedented opportunities for human coordination while exposing critical challenges in resource utilization efficiency. This technological landscape presents both remarkable achievements and fundamental limitations that demand innovative solutions.
	
	Social gaming and gamification achievements reveal remarkable progress in human engagement mechanisms. The evolution from early social networking platforms like Facebook (2004) and Twitter (2006) to sophisticated sandbox games including Minecraft (2011), Stardew Valley (2016), Terraria (2011), and Don't Starve (2013) has demonstrated the extraordinary potential of social connections and gamification mechanisms in driving sustained user engagement \cite{boyd2007social, persson2011minecraft, concernedape2016stardew, logic2011terraria, klei2013starve}. Location-based applications such as Pokémon GO (2016) achieved remarkable success by combining augmented reality, geolocation, and social mechanics, proving that mobile games could motivate real-world physical activity and social gathering on a global scale \cite{althoff2016influence, serino2016pokemon}. Environmental applications like Ant Forest (Alibaba's Alipay) have engaged over 600 million users in environmental activities, successfully translating virtual achievements into real-world reforestation efforts \cite{chen2020gamification}.
	
	Blockchain technology limitations present significant challenges despite technological advances. Since Bitcoin's introduction in 2008, blockchain technology has demonstrated that decentralized consensus is possible without central authorities, but achieves this at the cost of enormous energy consumption \cite{nakamoto2008bitcoin}. The Bitcoin network currently consumes approximately 150 TWh annually—equivalent to Argentina's electricity usage—performing calculations that contribute nothing beyond network security \cite{de2018bitcoin}. Even alternative consensus mechanisms like Ethereum's transition to Proof-of-Stake have reduced energy consumption but still fail to generate direct social value from the consensus process \cite{buterin2017casper, wood2014ethereum}.
	
	Blockchain social platform experiments have provided valuable but limited insights. Early blockchain social platform experiments including Steemit (2016), Minds (2015), Hive (2020), and LBRY/Odysee (2016) successfully demonstrated that decentralized incentive mechanisms could motivate user participation and content creation \cite{steemit2016whitepaper, minds2018whitepaper, hive2020constitution, kauffman2018lbry}. However, these pioneering platforms revealed fundamental limitations: subjective value assessment problems easily manipulated through vote buying and bot networks, plutocratic governance where wealthy participants dominate decision-making, lack of real-world impact verification, vulnerability to fraudulent manipulation, and unsustainable economic models leading to declining token values and user exodus \cite{xu2019analysis, hassan2019blockchain, renan2021hive, zhang2019gaming, liu2021tokenomics, wang2021sustainability}.
	
	Blockchain gaming evolution demonstrates both potential and sustainability challenges. The intersection of blockchain technology and gaming has produced valuable insights through platforms like CryptoKitties (2017), Axie Infinity (2018), The Sandbox (2020), Decentraland (2020), Gods Unchained (2019), and Splinterlands (2018) \cite{cryptokitties2017whitepaper, axieinfinity2020whitepaper, sandbox2020whitepaper, decentraland2017whitepaper, godsunchained2019whitepaper, splinterlands2020whitepaper}. These platforms demonstrated the powerful engagement potential of ownership, progression systems, and economic participation. However, most blockchain games struggle to maintain sustainable economies without continuous new player influx, leading to boom-bust cycles driven by speculation rather than genuine utility, creating scalability and user experience barriers, and facing regulatory uncertainties \cite{lehdonvirta2022play, murray2022speculation, kominers2022axie}.
	
	The convergence challenge reveals a fundamental technological paradox. This technological landscape presents a compelling situation: while social games and applications have proven exceptionally effective at motivating human engagement and behavior change, they typically lack mechanisms for creating lasting value beyond entertainment. Conversely, blockchain systems maintain valuable decentralized consensus properties but waste enormous resources on computationally intensive but socially meaningless tasks \cite{ball2017proofs}. Existing systems have failed to effectively integrate the engagement advantages of gamification with blockchain's potential for decentralized value creation.
	
	\subsection{EarthOL: Revolutionary Dual-Mainline Innovation Framework}
	
	To address these convergent challenges, EarthOL introduces a revolutionary dual-mainline gamification framework that organically integrates human civilization advancement with personal achievement development, creating unprecedented synergistic effects through systematic gamification of real-world contribution systems.
	
	\subsubsection{Human Civilization Technology Tree Mainline}
	
	Civilization-level progression systems treat the entirety of human civilization as a massive collaborative game, constructing multi-dimensional civilization technology trees spanning scientific technology, environmental protection, educational development, infrastructure construction, and cultural preservation. Each technology node represents milestones in humanity's collective knowledge and capabilities, requiring global user collaboration to unlock.
	
	Collective achievement mechanisms operate similar to guild systems in massive multiplayer online games, where users can choose specialization domains and contribute to civilization progress in those areas. For example, environmental technology tree nodes might include quantifiable civilization-level goals like "Global Forest Coverage Increase by 1
	
	Cross-cultural collaboration frameworks consider diverse cultural backgrounds and development levels, establishing inclusive contribution evaluation systems that ensure users from different regions and cultures can find appropriate contribution methods while avoiding technological colonialism tendencies.
	
	\subsubsection{Personal Life Achievement Mainline}
	
	Personalized development pathways provide each user with a unique personal achievement tree covering professional skills, social relationships, health levels, learning growth, and community contributions. The system intelligently recommends personalized achievement goals and development pathways based on users' interests, capabilities, and living environments.
	
	Skill certification and social value systems extend personal achievements beyond virtual rewards to integrate closely with real-world skill certification, career development, and social contributions. Skill certifications obtained through personal achievement completion can be used in real-world scenarios such as job seeking, education advancement, and social interaction, creating genuine social value.
	
	Life gamification experiences transform daily life activities including learning, working, exercising, and socializing through clear progress visualization, social comparison, and achievement celebration mechanisms, providing users with sustained satisfaction and motivation in pursuing personal development.
	
	\subsubsection{Dual-Mainline Synergy Mechanisms}
	
	Personal contribution amplification effects ensure that personal achievement completion not only advances individual development but simultaneously contributes progress toward corresponding civilization technology tree nodes. A user's personal achievements in education (such as obtaining teaching certification or significantly improving student outcomes) simultaneously advance the global education technology tree unlock process.
	
	Civilization feedback to personal development occurs when civilization technology tree unlocks open new achievement pathways and development opportunities for individual users. For example, when the "Quantum Computing Technology" civilization node is unlocked, users in related fields gain access to new personal skill learning pathways and career development opportunities.
	
	Social recognition and value realization through the dual-mainline mode ensure users experience both personal growth satisfaction and social value from contributing to human civilization progress, creating stronger sustained motivation than pure individual gaming or abstract social contribution.
	
	Cross-scale incentive mechanisms create sophisticated connections between individual behavior and collective goals, enabling EarthOL to establish an incentive system spanning personal, community, national, and global scales, where every user can achieve satisfaction at appropriate levels while driving broader social progress.
	
	\subsubsection{Core Innovation: Achievement-Based Proof-of-Work with Dual-Mainline Integration}
	
	The fundamental innovation underlying this dual-mainline framework is the Achievement-Based Proof-of-Work (APoW) system that replaces meaningless computational puzzles with verifiable personal achievements and contributions to human civilization advancement:
	
	\begin{equation}
		\text{APoW}_{score} = \sum_{domains} \text{ContributionValue}_{domain} \times \text{VerificationStrength} \times \text{CivilizationImpact}
	\end{equation}
	
	Where \text{CivilizationImpact} quantifies the long-term benefit to human society, moving beyond the self-referential value systems that plague existing blockchain social platforms and the speculative economics that characterize most blockchain games.
	
	Multi-layer decentralized verification protocols implement sophisticated verification architecture combining human judgment, algorithmic verification, and cross-cultural consensus to eliminate both centralized control and gaming vulnerabilities:
	
	\begin{algorithm}
		\caption{Contribution Authenticity Verification Protocol}
		\begin{algorithmic}
			\STATE \textbf{function} VerifyAuthenticity(contribution, validators):
			\STATE $pattern\_analysis \leftarrow$ AnalyzeBehavioralPatterns(validators)
			\STATE $social\_graph \leftarrow$ BuildSocialConnections(validators)
			\STATE $temporal\_correlation \leftarrow$ AnalyzeTimingPatterns(contribution, validators)
			\STATE $cross\_validation \leftarrow$ CompareWithExternalSources(contribution)
			\STATE $impact\_verification \leftarrow$ VerifyRealWorldImpact(contribution)
			\STATE 
			\STATE $authenticity\_score \leftarrow$ WeightedCombination(pattern\_analysis, social\_graph, temporal\_correlation, cross\_validation, impact\_verification)
			\STATE 
			\STATE \textbf{if} $authenticity\_score < THRESHOLD$ \textbf{then}
			\STATE \quad TriggerInvestigation(contribution, validators)
			\STATE \quad ProtectSystemIntegrity(suspected\_fraud)
			\STATE \textbf{else}
			\STATE \quad CelebrateAuthenticContribution(contribution)
			\STATE \quad RewardGameplayExperience(contributors, validators)
			\STATE \textbf{end if}
		\end{algorithmic}
	\end{algorithm}
	
	Fortress-level security architecture implements military-grade security measures with economic incentives for security specialists:
	
	\begin{equation}
		\text{SecurityReward} = \text{BaseReward} \times \text{ThreatLevel} \times \text{PreventionEffectiveness} \times \text{ResponseTime}^{-1}
	\end{equation}
	
	\subsubsection{Domain-Restricted Value Assessment with Objective Impact Measurement}
	
	The core innovation of EarthOL lies in addressing the "value assessment problem" through domain-restricted evaluation rather than attempting universal value assessment. We categorize domains by their suitability for human contribution verification using a Domain Feasibility Framework with Impact Weighting:
	
	\begin{alignat}{2}
		&\text{High Feasibility } (F > 0.7) &&: \,
		\begin{minipage}[t]{0.65\linewidth}
			\footnotesize\hangindent=1em Environmental data collection with satellite verification, educational content with learning outcome measurement, community problem solving with documented impact
		\end{minipage} \\
		&\text{Medium Feasibility } (0.4 < F < 0.7) &&: \,
		\begin{minipage}[t]{0.65\linewidth}
			\footnotesize\hangindent=1em Local infrastructure assessment with usage metrics, cultural preservation with engagement tracking, health awareness with behavior change measurement
		\end{minipage} \\
		&\text{Low Feasibility } (F < 0.4) &&: \,
		\begin{minipage}[t]{0.65\linewidth}
			\footnotesize\hangindent=1em Pure artistic expression, political advocacy, personal relationship advice
		\end{minipage}
	\end{alignat}
	
	Where $F$ represents a composite score of objectivity, verifiability, cultural consensus, and measurable civilization impact—transcending the simple engagement metrics used by existing blockchain social platforms and the speculative value metrics that characterize blockchain gaming.
	
	\subsubsection{The EarthOL Vision: Meaningful, Enjoyable, and Impactful Life}
	
	The dual-mainline mode's core innovation lies in solving existing blockchain social platforms' value assessment difficulties and sustainability problems while creating a complete incentive ecosystem with both intrinsic drive and external social significance through deep integration of personal development and civilization progress. Users naturally contribute to human civilization advancement while pursuing personal achievements, and civilization progress provides broader stages and richer opportunities for personal development.
	
	EarthOL's ultimate goal is not merely creating another blockchain system, but fundamentally transforming how humans experience work, contribution, and achievement in the modern world. By treating real-world civilization advancement as an engaging, collaborative game with clear progression systems, meaningful achievements, and social recognition, we aim to make every person's daily activities more enjoyable while simultaneously accelerating human progress.
	
	The subsequent sections present detailed theoretical foundations, implementation architecture, and empirical analysis demonstrating how the convergence of gaming excellence and blockchain innovation can be harnessed to create meaningful human-centered consensus systems within carefully bounded domains, supported by the revolutionary dual-mainline framework that makes life both personally fulfilling and civilizationally impactful.
	
	\section{Theoretical Foundations and Limitations}
	
	\subsection{The Value Assessment Problem}
	
	\subsubsection{Arrow's Impossibility and Domain Restriction}
	
	Arrow's impossibility theorem proves that no voting system can simultaneously satisfy basic fairness criteria when aggregating diverse preferences. This creates fundamental constraints for any system attempting to assess contribution value democratically.
	
	\textbf{Our Response - Domain Restriction}: We restrict the domain of possible preferences to those with sufficient cultural consensus:
	
	\begin{equation}
		\mathcal{D}_{viable} = \{P \in \mathcal{D} : \text{CrossCulturalConsensus}(P) > \theta_{min}\}
	\end{equation}
	
	Where $\theta_{min} = 0.6$ represents the minimum consensus threshold for viable domains.
	
	\subsubsection{Quantifying Domain Feasibility}
	
	\textbf{Feasibility Score Calculation}:
	\begin{equation}
		F_{domain} = 0.4 \cdot \text{Objectivity} + 0.3 \cdot \text{Verifiability} + 0.3 \cdot \text{CulturalConsensus}
	\end{equation}
	
	\textbf{Component Definitions}:
	\begin{align}
		\text{Objectivity} &= 1 - \frac{\text{Var(CrossCulturalAssessments)}}{\text{Mean(CrossCulturalAssessments)}}\\
		\text{Verifiability} &= \frac{\text{VerifiableAspects}}{\text{TotalAspects}}\\
		\text{CulturalConsensus} &= 1 - \text{KL}_{\text{divergence}}(\text{Culture}_A || \text{Culture}_B)
	\end{align}
	
	\textbf{Empirical Domain Rankings} (based on theoretical analysis):
	\begin{itemize}
		\item \textbf{Open Source Software}: $F = 0.82$ (high algorithmic verification + clear impact metrics)
		\item \textbf{Mathematical Proofs}: $F = 0.79$ (universal logical standards)
		\item \textbf{Data Analysis}: $F = 0.76$ (reproducible results + statistical validation)
		\item \textbf{Scientific Research}: $F = 0.65$ (peer review traditions but methodology debates)
		\item \textbf{Educational Content}: $F = 0.58$ (learning outcomes measurable but pedagogy varies)
		\item \textbf{Artistic Expression}: $F = 0.28$ (highly subjective aesthetic judgments)
	\end{itemize}
	
	\subsection{Game-Theoretic Analysis}
	
	\subsubsection{Strategic Contribution Behavior}
	
	Participants will optimize their contribution strategies to maximize rewards rather than genuine social value.
	
	\textbf{Revised Validator Utility Function}:
	\begin{equation}
		U_{validator}(effort, accuracy) = \beta \cdot \text{Reward}(accuracy) - \text{Cost}(effort) + \gamma \cdot \text{Reputation}(\Delta)
	\end{equation}
	
	\textbf{Nash Equilibrium Conditions} (simplified):
	For honest validation to be a Nash equilibrium:
	\begin{equation}
		\beta \cdot \frac{\partial \text{Reward}}{\partial accuracy} \geq \frac{\partial \text{Cost}}{\partial effort}
	\end{equation}
	
	\textbf{Critical Parameter Bounds} (from stability analysis):
	\begin{itemize}
		\item Detection probability threshold: $P(\text{detection}) > 0.3$
		\item Penalty-to-reward ratio: $\frac{\text{Penalty}}{\text{Reward}} > 5$
		\item Reputation decay factor: $\lambda \in [0.90, 0.95]$
	\end{itemize}
	
	\subsubsection{Simplified Collusion Resistance}
	
	\textbf{Collusion Detection Metric}:
	\begin{equation}
		\text{CollusionScore} = \frac{\text{AgreementRate} \times \text{SocialProximity}}{\text{ExpectedRandomRate}}
	\end{equation}
	
	\textbf{Byzantine Fault Tolerance}: System remains secure with up to $\lfloor n/3 \rfloor$ colluding validators, provided:
	\begin{itemize}
		\item Random validator assignment for each contribution
		\item Multi-round validation with different validator sets
		\item Reputation-weighted voting with quadratic penalties
	\end{itemize}
	
	\subsection{Cultural Bias Mitigation}
	
	\textbf{Bias Minimization Strategy}:
	Rather than attempting to eliminate bias (impossible), we minimize it through:
	
	\begin{enumerate}
		\item \textbf{Regional Validator Matching}: Contributors matched with culturally similar validators for initial assessment
		\item \textbf{Cross-Cultural Verification}: High-value contributions require validation from multiple cultural perspectives
		\item \textbf{Explicit Bias Measurement}: Quantitative tracking of cultural bias in validation outcomes
	\end{enumerate}
	
	\textbf{Bias Lower Bound} (theoretical result):
	For any non-trivial multi-cultural system:
	\begin{equation}
		\text{Bias}_{min} \geq \max_{cultures} \text{KL}(P_i || P_j) \cdot P(\text{conflict})
	\end{equation}
	
	Where $P(\text{conflict})$ is the probability of value disagreement between cultures $i$ and $j$.
	
	\section{Enhanced Multi-Domain Protocol Architecture with Advanced Security}
	
	\subsection{Comprehensive Five-Layer Verification System}
	
	Based on feasibility analysis and security requirements, we implement an enhanced verification architecture with comprehensive security measures:
	
	\begin{figure}[ht]
		\centering
		\begin{tikzpicture}[
			box/.style={rectangle, draw, fill=blue!20, minimum height=1cm, minimum width=3cm, text centered, text width=3cm},
			arrow/.style={->, thick},
			security/.style={rectangle, draw, fill=red!20, minimum height=0.8cm, minimum width=2.5cm, text centered, text width=2.5cm},
			validator/.style={rectangle, draw, fill=green!20, minimum height=0.8cm, minimum width=2.5cm, text centered, text width=2.5cm}
			]
			
			\node[box] (layer1) at (0,0) {Layer 1: Algorithmic Pre-Screening};
			\node[box] (layer2) at (0,-2) {Layer 2: Community Validation};
			\node[box] (layer3) at (0,-4) {Layer 3: Expert Review};
			\node[box] (layer4) at (0,-6) {Layer 4: Cross-Cultural Consensus};
			\node[box] (layer5) at (0,-8) {Layer 5: Long-term Impact Assessment};
			
			\node[security] (sec1) at (4,1) {Authenticity Detection};
			\node[security] (sec2) at (4,-1) {Collusion Monitoring};
			\node[security] (sec3) at (4,-3) {Identity Verification};
			\node[security] (sec4) at (4,-5) {Bias Detection};
			\node[security] (sec5) at (4,-7) {Fraud Prevention};
			
			\node[validator] (val1) at (8,0) {Algorithmic Validators};
			\node[validator] (val2) at (8,-2) {Community Validators};
			\node[validator] (val3) at (8,-4) {Expert Validators};
			\node[validator] (val4) at (8,-6) {Cultural Ambassadors};
			\node[validator] (val5) at (8,-8) {Impact Assessors};
			
			\draw[arrow] (layer1) -- (layer2);
			\draw[arrow] (layer2) -- (layer3);
			\draw[arrow] (layer3) -- (layer4);
			\draw[arrow] (layer4) -- (layer5);
			
			\draw[arrow, dashed, red] (sec1) -- (layer1);
			\draw[arrow, dashed, red] (sec2) -- (layer2);
			\draw[arrow, dashed, red] (sec3) -- (layer3);
			\draw[arrow, dashed, red] (sec4) -- (layer4);
			\draw[arrow, dashed, red] (sec5) -- (layer5);
			
			\draw[arrow, dashed, green] (val1) -- (layer1);
			\draw[arrow, dashed, green] (val2) -- (layer2);
			\draw[arrow, dashed, green] (val3) -- (layer3);
			\draw[arrow, dashed, green] (val4) -- (layer4);
			\draw[arrow, dashed, green] (val5) -- (layer5);
			
		\end{tikzpicture}
		\caption{Enhanced Five-Layer Verification Architecture with Security and Validator Incentive Systems}
	\end{figure}
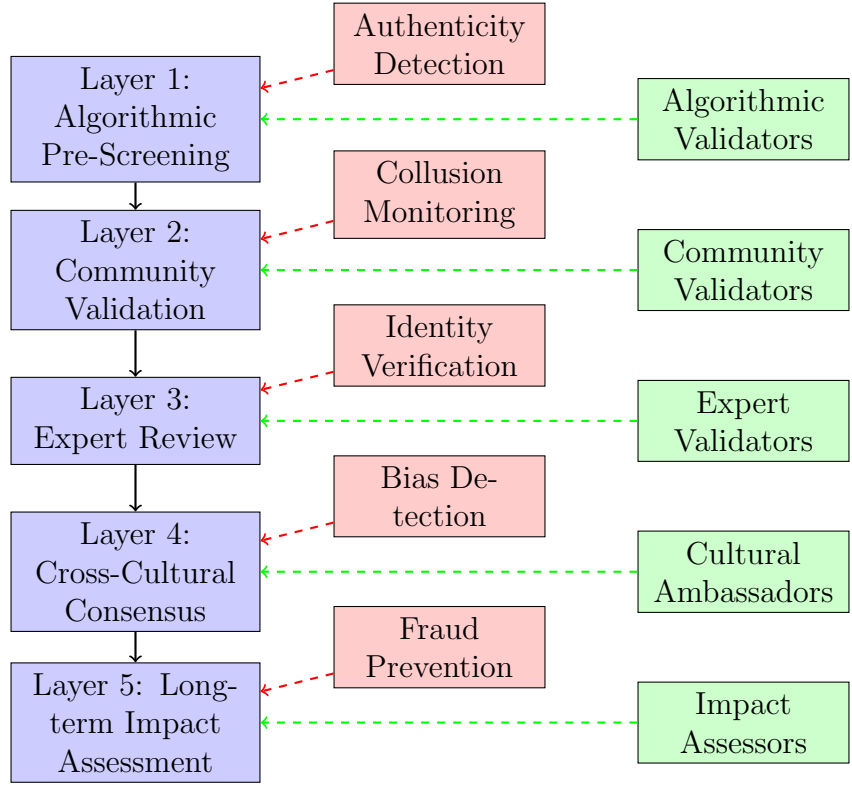
	
	\subsubsection{Layer 1: Advanced Algorithmic Pre-Screening}
	\textbf{Target Domains}: Code contributions, mathematical proofs, data analysis\\
	\textbf{Verification Method}: Multi-algorithm ensemble with security checks\\
	\textbf{Capacity}: 2000+ contributions/day\\
	\textbf{Accuracy}: 90-97\%
	
	\textbf{Enhanced Core Algorithm}:
	\begin{algorithm}
		\caption{Enhanced Layer 1 Verification with Security}
		\begin{algorithmic}
			\STATE \textbf{function} Layer1VerifyEnhanced(contribution):
			\STATE $correctness \leftarrow$ RunAutomatedTests(contribution)
			\STATE $novelty \leftarrow$ CheckSimilarity(contribution, database)
			\STATE $security \leftarrow$ DetectMaliciousPatterns(contribution)
			\STATE $authenticity \leftarrow$ VerifyAuthorshipSignature(contribution)
			\STATE $complexity \leftarrow$ AnalyzeComputationalComplexity(contribution)
			\STATE 
			\STATE \textbf{if} $security < 0.8$ \textbf{then}
			\STATE \quad FlagForSecurityReview(contribution)
			\STATE \quad \textbf{return} SECURITY\_FLAGGED
			\STATE \textbf{end if}
			\STATE 
			\STATE $base\_score \leftarrow 0.4 \cdot correctness + 0.3 \cdot novelty + 0.2 \cdot complexity + 0.1 \cdot authenticity$
			\STATE $final\_score \leftarrow base\_score \cdot security$
			\STATE 
			\STATE \textbf{if} $final\_score > 0.85$ \textbf{then}
			\STATE \quad RewardAlgorithmicValidator(validator\_id, TIER\_1\_REWARD)
			\STATE \textbf{end if}
			\STATE 
			\STATE \textbf{return} $final\_score$
		\end{algorithmic}
	\end{algorithm}
	
	\textbf{Algorithmic Validator Incentives}:
	\begin{itemize}
		\item Base reward: 10-50 tokens per validated contribution
		\item Accuracy bonus: Up to 100\% bonus for 95\%+ accuracy
		\item Speed bonus: 25\% bonus for sub-second processing
		\item Security detection bonus: 200\% bonus for identifying malicious content
		\item Achievement system: "Lightning Validator", "Security Sentinel", "Accuracy Master"
	\end{itemize}
	
	\subsubsection{Layer 2: Enhanced Community Validation with Gamification}
	\textbf{Target Domains}: Educational content, technical documentation, community projects\\
	\textbf{Verification Method}: Structured community voting with bias correction and gaming elements\\
	\textbf{Capacity}: 800-1200 contributions/day\\
	\textbf{Accuracy}: 75-88\%
	
	\textbf{Enhanced Bias-Corrected Scoring with Security}:
	\begin{equation}
		\text{Score}(C) = \sum_i w_i \cdot \text{Vote}_i(C) \cdot (1 + \text{BiasCorrection}_i) \cdot \text{SecurityWeight}_i
	\end{equation}
	
	Where:
	\begin{align}
		\text{SecurityWeight}_i &= \begin{cases}
			1.0 & \text{if validator verified and trusted} \\
			0.5 & \text{if validator new or flagged} \\
			0.0 & \text{if validator suspected of fraud}
		\end{cases}
	\end{align}
	
	\textbf{Community Validator Gamification System}:
	\begin{itemize}
		\item \textbf{Base Rewards}: 5-25 tokens per validation
		\item \textbf{Consensus Bonus}: Extra rewards for agreeing with final consensus
		\item \textbf{Quality Recognition}: "Quality Curator", "Community Champion" badges
		\item \textbf{Leaderboards}: Weekly/monthly top validator rankings
		\item \textbf{Social Features}: Validator profiles, contribution galleries, peer endorsements
		\item \textbf{Progressive Unlocks}: Access to higher-tier validation tasks with experience
	\end{itemize}
	
	\textbf{Authenticity Protection Measures for Community Layer}:
	\begin{enumerate}
		\item \textbf{Temporal Validation Patterns}: Monitor for suspicious voting patterns
		\item \textbf{Social Graph Analysis}: Detect coordinated behavior among validators
		\item \textbf{Quality Consistency Tracking}: Flag validators with inconsistent accuracy
		\item \textbf{Multi-Identity Detection}: Prevent sybil attacks through behavioral analysis
	\end{enumerate}
	
	\subsubsection{Layer 3: Expert Review with Professional Incentives}
	\textbf{Target Domains}: Scientific research, technical innovation, complex analysis\\
	\textbf{Verification Method}: Domain expert evaluation with reputation weighting and peer review\\
	\textbf{Capacity}: 50-150 contributions/day\\
	\textbf{Accuracy}: 92-99\%
	
	\textbf{Enhanced Expert Selection Algorithm}:
	\begin{equation}
		P(\text{select}|\text{expert}_i) = \frac{\text{Expertise}_i \cdot \text{Reputation}_i \cdot \text{Availability}_i \cdot \text{DomainMatch}_i}{\sum_j (\text{Expertise}_j \cdot \text{Reputation}_j \cdot \text{Availability}_j \cdot \text{DomainMatch}_j)}
	\end{equation}
	
	\textbf{Expert Validator Incentive System}:
	\begin{itemize}
		\item \textbf{Premium Rewards}: 100-500 tokens per expert review
		\item \textbf{Professional Recognition}: Verified expert badges, institutional affiliations
		\item \textbf{Career Benefits}: Publication credits, peer network expansion
		\item \textbf{Impact Tracking}: Citations and downstream effects of validated work
		\item \textbf{Exclusive Access}: First access to cutting-edge contributions in their domain
		\item \textbf{Mentorship Opportunities}: Pairing with junior validators for training
	\end{itemize}
	
	\subsubsection{Layer 4: Cross-Cultural Consensus with Cultural Ambassador Program}
	\textbf{Target Domains}: Culturally sensitive contributions, global impact projects\\
	\textbf{Verification Method}: Multi-cultural validator panels with cultural sensitivity algorithms\\
	\textbf{Capacity}: 20-80 contributions/day\\
	\textbf{Accuracy}: 70-85\% (varies by cultural complexity)
	
	\textbf{Cultural Ambassador Incentives}:
	\begin{itemize}
		\item \textbf{Cultural Bridge Rewards}: 50-200 tokens for successful cross-cultural consensus
		\item \textbf{Diplomatic Achievements}: "Cultural Bridge", "Global Harmony" badges
		\item \textbf{Language Bonuses}: Extra rewards for multilingual validation
		\item \textbf{Cultural Education}: Access to cultural competency training modules
		\item \textbf{International Recognition}: Global leaderboards, international validator exchanges
	\end{itemize}
	
	\subsubsection{Layer 5: Long-term Impact Assessment with Impact Specialists}
	\textbf{Target Domains}: All domains requiring longitudinal validation\\
	\textbf{Verification Method}: Time-series analysis of contribution outcomes\\
	\textbf{Capacity}: 10-30 contributions/day\\
	\textbf{Accuracy}: 80-95\% (measured over 6-24 month periods)
	
	\textbf{Impact Assessor Incentives}:
	\begin{itemize}
		\item \textbf{Long-term Rewards}: Delayed but substantial rewards (200-1000 tokens)
		\item \textbf{Impact Multipliers}: Rewards scale with measured real-world impact
		\item \textbf{Research Opportunities}: Access to longitudinal data for academic research
		\item \textbf{Policy Influence}: Input into platform governance and parameter tuning
		\item \textbf{Legacy Tracking}: Personal impact portfolios showing long-term contributions
	\end{itemize}
	
	\subsection{Advanced Security Framework}
	
	\subsubsection{Multi-Layer Security Architecture}
	
	\begin{figure}[ht]
		\centering
		\begin{tikzpicture}[
			layer/.style={rectangle, draw, fill=blue!10, minimum height=2.2cm, minimum width=16.1cm, text centered},
			threat/.style={ellipse, draw, fill=red!20, minimum height=0.3cm, minimum width=2cm, text centered, text width=3.5cm},
			defense/.style={rectangle, draw, fill=green!20, minimum height=0.6cm, minimum width=2cm, text centered, text width=4.2cm}
			]
			
			\node[layer] (crypto) at (0,0) {Cryptographic Security Layer};
			\node[layer] (network) at (0,-2.5) {Network Security Layer};
			\node[layer] (behavioral) at (0,-5) {Behavioral Security Layer};
			\node[layer] (social) at (0,-7.5) {Social Security Layer};
			
			\node[threat] (t1a) at (-5.3,0.5) {Key Compromise};
			\node[threat] (t1b) at (-5.3,-0.5) {Signature Forgery};
			\node[defense] (d1a) at (5.3,0.5) {Multi-Sig Validation};
			\node[defense] (d1b) at (5.3,-0.5) {Zero-Knowledge Proofs};
			
			\node[threat] (t2a) at (-5.3,-2) {Sybil Attacks};
			\node[threat] (t2b) at (-5.3,-3) {DDoS Attacks};
			\node[defense] (d2a) at (5.3,-2) {Identity Verification};
			\node[defense] (d2b) at (5.3,-3) {Rate Limiting};
			
			\node[threat] (t3a) at (-5.3,-4.5) {Fraudulent Behavior};
			\node[threat] (t3b) at (-5.3,-5.5) {Pattern Abuse};
			\node[defense] (d3a) at (5.3,-4.5) {ML-Based Detection};
			\node[defense] (d3b) at (5.3,-5.5) {Anomaly Scoring};
			
			\node[threat] (t4a) at (-5.3,-7) {Collusion Networks};
			\node[threat] (t4b) at (-5.3,-8) {Social Engineering};
			\node[defense] (d4a) at (5.3,-7) {Graph Analysis};
			\node[defense] (d4b) at (5.3,-8) {Trust Scoring};
			
			\draw[<->, red, thick] (t1a) -- (d1a);
			\draw[<->, red, thick] (t1b) -- (d1b);
			\draw[<->, red, thick] (t2a) -- (d2a);
			\draw[<->, red, thick] (t2b) -- (d2b);
			\draw[<->, red, thick] (t3a) -- (d3a);
			\draw[<->, red, thick] (t3b) -- (d3b);
			\draw[<->, red, thick] (t4a) -- (d4a);
			\draw[<->, red, thick] (t4b) -- (d4b);
			
		\end{tikzpicture}
		\caption{Multi-Layer Security Architecture with Threat-Defense Mapping}
	\end{figure}
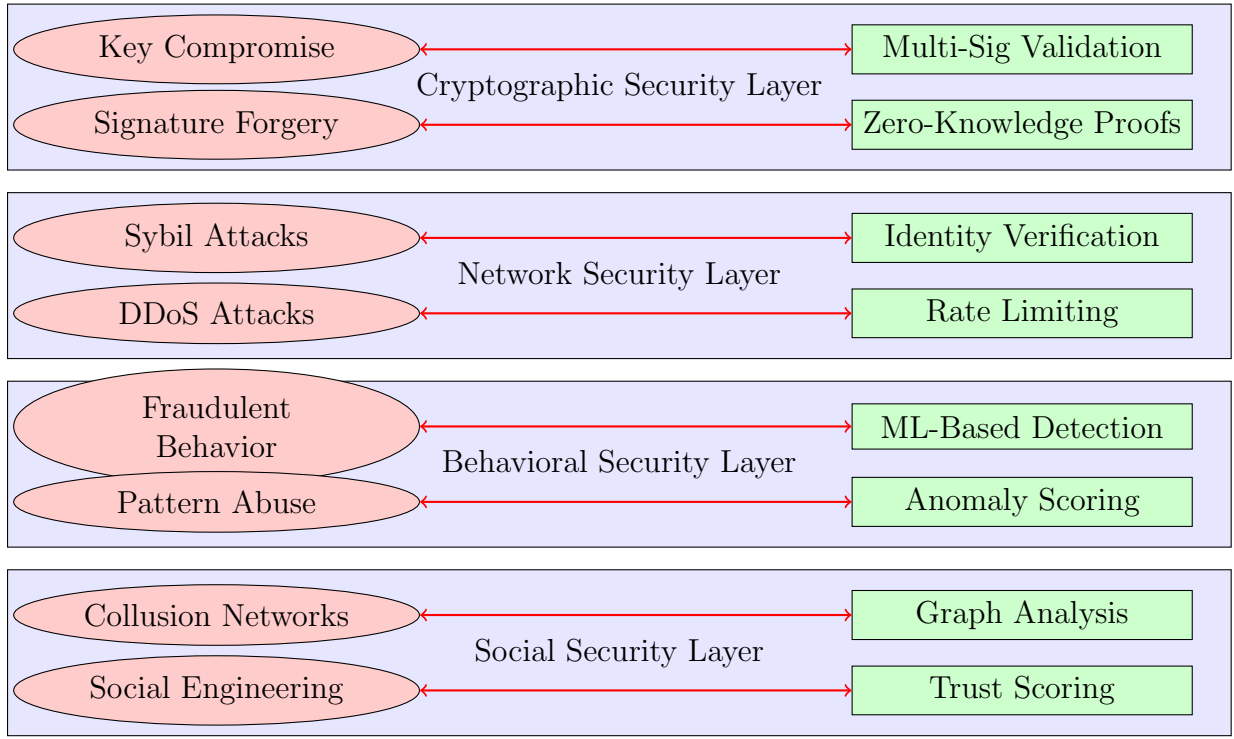
	
	\subsubsection{Cryptographic Security Layer}
	
	\textbf{Enhanced Digital Signature System}:
	\begin{equation}
		\text{Signature}_{enhanced} = \text{Sign}_{private}(\text{Hash}(\text{contribution} || \text{timestamp} || \text{metadata}))
	\end{equation}
	
	\textbf{Multi-Signature Validation Requirements}:
	\begin{itemize}
		\item Contributor signature (required)
		\item Validator signatures (minimum 3 for Layer 2+)
		\item System timestamp signature (automatic)
		\item Domain expert signature (for Layer 3+)
	\end{itemize}
	
	\textbf{Zero-Knowledge Proof Integration}:
	For sensitive contributions, validators can verify validity without accessing full content:
	\begin{equation}
		\text{ZK-Proof} = \text{Prove}(\text{ValidContribution}(\text{hidden\_content}), \text{public\_criteria})
	\end{equation}
	
	\subsubsection{Network Security Layer}
	
	\textbf{Advanced Sybil Attack Prevention}:
	\begin{algorithm}
		\caption{Enhanced Sybil Detection}
		\begin{algorithmic}
			\STATE \textbf{function} DetectSybilAttack(validator\_set):
			\STATE $identity\_features \leftarrow$ ExtractIdentityFeatures(validator\_set)
			\STATE $behavioral\_patterns \leftarrow$ AnalyzeBehavioralPatterns(validator\_set)
			\STATE $social\_graph \leftarrow$ BuildSocialConnectionGraph(validator\_set)
			\STATE 
			\STATE $similarity\_matrix \leftarrow$ ComputeSimilarityMatrix(identity\_features, behavioral\_patterns)
			\STATE $clusters \leftarrow$ DetectSuspiciousClusters(similarity\_matrix, social\_graph)
			\STATE 
			\STATE \textbf{for each} cluster \textbf{in} clusters \textbf{do}
			\STATE \quad $sybil\_score \leftarrow$ ComputeSybilScore(cluster)
			\STATE \quad \textbf{if} $sybil\_score > SYBIL\_THRESHOLD$ \textbf{then}
			\STATE \quad \quad FlagPotentialSybils(cluster)
			\STATE \quad \quad RewardSecuritySpecialist(detector\_id, SECURITY\_REWARD)
			\STATE \quad \textbf{end if}
			\STATE \textbf{end for}
		\end{algorithmic}
	\end{algorithm}
	
	\textbf{DDoS Protection and Rate Limiting}:
	\begin{itemize}
		\item Progressive rate limiting based on reputation
		\item Proof-of-work requirements for new accounts
		\item Geographic distribution of validation nodes
		\item Adaptive threshold adjustment during attacks
	\end{itemize}
	
	\subsubsection{Behavioral Security Layer}
	
	\textbf{Machine Learning-Based Fraud Detection}:
	\begin{equation}
		\text{Fraud\_Score} = \text{ML\_Model}(\text{validation\_patterns}, \text{timing\_analysis}, \text{quality\_consistency})
	\end{equation}
	
	\textbf{Anomaly Detection Features}:
	\begin{itemize}
		\item Validation timing patterns
		\item Quality score distributions
		\item Social interaction patterns
		\item Contribution-validator matching anomalies
	\end{itemize}
	
	\textbf{Security Specialist Incentives}:
	\begin{itemize}
		\item \textbf{Detection Rewards}: 100-500 tokens per confirmed fraud detection
		\item \textbf{Security Achievements}: "Guardian Angel", "System Defender" badges
		\item \textbf{Research Access}: Access to anonymized attack data for security research
		\item \textbf{Bug Bounties}: Rewards for discovering new attack vectors
		\item \textbf{Security Leaderboards}: Recognition for top security contributors
	\end{itemize}
	
	\subsection{Enhanced Token System with Validator Rewards}
	
	\subsubsection{Multi-Tier Token Economy}
	
	Based on comprehensive feedback analysis, we implement a sophisticated multi-token system with role-specific incentives:
	
	\textbf{Primary Token Issuance}:
	\begin{equation}
		\text{NewTokens} = \text{BaseReward} \times \text{QualityMultiplier} \times \text{DomainWeight} \times \text{AntiInflationFactor} \times \text{SecurityBonus}
	\end{equation}
	
	\textbf{Validator-Specific Token Allocation}:
	\begin{align}
		\text{ContributorTokens} &= 0.60 \times \text{NewTokens}\\
		\text{ValidatorTokens} &= 0.25 \times \text{NewTokens}\\
		\text{SecurityTokens} &= 0.10 \times \text{NewTokens}\\
		\text{SystemReserve} &= 0.05 \times \text{NewTokens}
	\end{align}
	
	\textbf{Enhanced Domain Weights} (to balance different contribution types):
	\begin{itemize}
		\item Technical Implementation: 1.2
		\item Scientific Research: 1.1
		\item Educational Content: 1.0
		\item Community Projects: 0.9
		\item Cultural Preservation: 0.8
		\item Creative Expression: 0.7
	\end{itemize}
	
	\subsubsection{Validator Achievement and Ranking Systems}
	
	\textbf{Achievement Categories}:
	\begin{enumerate}
		\item \textbf{Accuracy Achievements}:
		\begin{itemize}
			\item Precision Master (95\%+ accuracy over 100 validations)
			\item Consistency Champion (maintaining accuracy for 6+ months)
			\item Perfect Streak (50+ consecutive accurate validations)
		\end{itemize}
		
		\item \textbf{Volume Achievements}:
		\begin{itemize}
			\item Validation Veteran (1000+ validations completed)
			\item Daily Dedicated (validation activity every day for 30+ days)
			\item Speed Demon (fastest average validation time in category)
		\end{itemize}
		
		\item \textbf{Security Achievements}:
		\begin{itemize}
			\item Threat Hunter (detecting 10+ fraud attempts)
			\item System Guardian (preventing major security incidents)
			\item Vulnerability Discoverer (finding new attack vectors)
		\end{itemize}
		
		\item \textbf{Social Achievements}:
		\begin{itemize}
			\item Mentor Master (training 50+ new validators)
			\item Cross-Cultural Bridge (successful multi-cultural validations)
			\item Community Builder (organizing validator events)
		\end{itemize}
	\end{enumerate}
	
	\textbf{Dynamic Ranking System}:
	\begin{equation}
		\text{ValidatorRank} = \sum_{categories} w_i \times \text{Score}_i \times \text{RecencyFactor}_i \times \text{DifficultyMultiplier}_i
	\end{equation}
	
	Where:
	\begin{align}
		\text{RecencyFactor}_i &= e^{-\lambda \times \text{DaysSinceLastActivity}}\\
		\text{DifficultyMultiplier}_i &= 1 + 0.1 \times \text{AverageContributionComplexity}
	\end{align}
	
	\subsection{Enhanced Authenticity Protection Mechanisms}
	
	\subsubsection{Multi-Evidence Verification Protocol}
	
	Each contribution must provide comprehensive evidence across multiple verification channels:
	
	\begin{figure}[ht]
		\centering
		\begin{tikzpicture}[
			evidence/.style={rectangle, draw, fill=yellow!20, minimum height=1cm, minimum width=2.5cm, text centered, text width=2.5cm},
			verification/.style={diamond, draw, fill=blue!20, minimum height=1cm, minimum width=1.5cm, text centered, aspect=2, text width=2cm},
			result/.style={ellipse, draw, fill=green!20, minimum height=0.8cm, minimum width=2cm, text centered, text width=2cm}
			]
			
			\node[evidence] (temporal) at (-5,3) {Temporal Evidence};
			\node[evidence] (social) at (-9,0) {Social Evidence};
			\node[evidence] (technical) at (-6,-2) {Technical Evidence};
			\node[evidence] (impact) at (-3,-3) {Impact Evidence};
			
			\node[verification] (verify) at (0,0) {Multi-Evidence Verification};
			
			\node[result] (authentic) at (4,2) {Authentic Contribution};
			\node[result] (suspicious) at (4,-2) {Suspicious Contribution};
			
			\draw[->] (temporal) -- (verify);
			\draw[->] (social) -- (verify);
			\draw[->] (technical) -- (verify);
			\draw[->] (impact) -- (verify);
			
			\draw[->] (verify) -- (authentic);
			\draw[->] (verify) -- (suspicious);
			
			\node[above] at (-2.5,1.5) {Timeline Analysis};
			\node[above] at (-5.3,0) {Community Validation};
			\node[below] at (-4,-0.5) {Code/Data Verification};
			\node[below] at (-1.5,-1.5) {Usage Metrics};
			
		\end{tikzpicture}
		\caption{Multi-Evidence Verification Protocol}
	\end{figure}
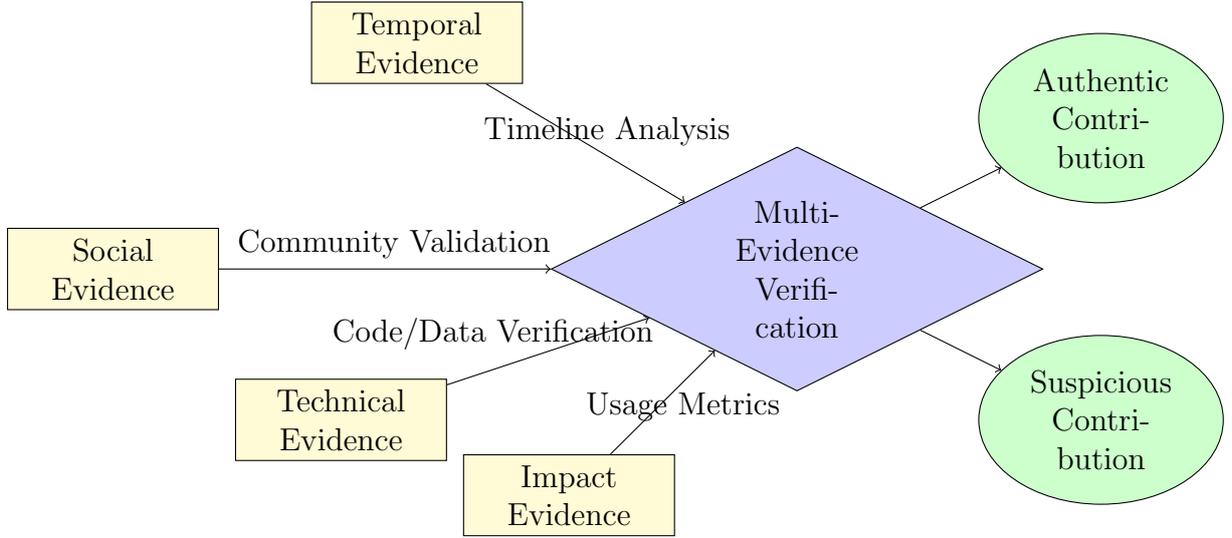
	
	\textbf{Evidence Requirements by Layer}:
	\begin{enumerate}
		\item \textbf{Temporal Evidence}: Development timeline with cryptographic timestamps
		\item \textbf{Social Evidence}: Community acknowledgment, collaboration records, peer endorsements
		\item \textbf{Technical Evidence}: Reproducible results, executable code, detailed methodology
		\item \textbf{Impact Evidence}: Usage statistics, adoption metrics, measurable outcomes
	\end{enumerate}
	
	\subsubsection{Advanced Reputation System with Multi-Dimensional Scoring}
	
	\textbf{Enhanced Reputation Formula}:
	\begin{equation}
		\text{Rep}_{t+1} = \alpha \cdot \text{Rep}_t + (1-\alpha) \cdot \text{WeightedCurrentPerformance}
	\end{equation}
	
	Where:
	\begin{align}
		\text{WeightedCurrentPerformance} &= \sum_{dimensions} w_i \times \text{Performance}_i\\
		\text{Dimensions} &= \{\text{Accuracy, Speed, Security, Consistency, Innovation}\}
	\end{align}
	
	\textbf{Reputation-Based Validation Weights with Security Adjustments}:
	\begin{equation}
		\text{Weight}(\text{validator}) = \frac{\text{Rep}(\text{validator})^2 \times \text{SecurityScore}(\text{validator})}{\sum_{all} \text{Rep}(\text{validator})^2 \times \text{SecurityScore}(\text{validator})}
	\end{equation}
	
	\subsubsection{Collusion Detection and Prevention}
	
	\textbf{Advanced Collusion Detection Algorithm}:
	\begin{algorithm}
		\caption{Multi-Dimensional Collusion Detection}
		\begin{algorithmic}
			\STATE \textbf{function} DetectCollusion(validator\_network):
			\STATE $social\_graph \leftarrow$ BuildSocialConnectionGraph(validator\_network)
			\STATE $validation\_patterns \leftarrow$ AnalyzeValidationPatterns(validator\_network)
			\STATE $timing\_correlations \leftarrow$ ComputeTimingCorrelations(validator\_network)
			\STATE $quality\_correlations \leftarrow$ ComputeQualityCorrelations(validator\_network)
			\STATE 
			\STATE \textbf{for each} validator\_pair \textbf{in} validator\_network \textbf{do}
			\STATE \quad $social\_proximity \leftarrow$ ComputeSocialProximity(validator\_pair, social\_graph)
			\STATE \quad $pattern\_similarity \leftarrow$ ComputePatternSimilarity(validator\_pair, validation\_patterns)
			\STATE \quad $timing\_correlation \leftarrow$ GetTimingCorrelation(validator\_pair, timing\_correlations)
			\STATE \quad $quality\_correlation \leftarrow$ GetQualityCorrelation(validator\_pair, quality\_correlations)
			\STATE \quad 
			\STATE \quad $collusion\_score \leftarrow$ CombineCollusionSignals(social\_proximity, pattern\_similarity, timing\_correlation, quality\_correlation)
			\STATE \quad 
			\STATE \quad \textbf{if} $collusion\_score > COLLUSION\_THRESHOLD$ \textbf{then}
			\STATE \quad \quad FlagPotentialCollusion(validator\_pair)
			\STATE \quad \quad TriggerInvestigation(validator\_pair)
			\STATE \quad \quad RewardCollusionDetector(detector\_id, COLLUSION\_DETECTION\_REWARD)
			\STATE \quad \textbf{end if}
			\STATE \textbf{end for}
		\end{algorithmic}
	\end{algorithm}
	
	\section{Probabilistic Modeling and Enhanced Scalability Analysis}
	
	\subsection{Advanced System Performance Modeling}
	
	\subsubsection{Enhanced Throughput Analysis Using Advanced Queueing Theory}
	
	\textbf{Multi-Class M/M/c Queue Model} for the enhanced validation system:
	- Arrival rates: $\lambda_i$ contributions/hour for layer $i$
	- Service rates: $\mu_{ij}$ validations/hour per validator type $j$ in layer $i$  
	- Number of validators: $c_{ij}$ validators of type $j$ in layer $i$
	- Priority weights: $w_i$ for layer $i$
	
	\textbf{Enhanced Key Performance Metrics}:
	\begin{align}
		\text{System utilization:} \quad \rho_{ij} &= \frac{\lambda_i}{c_{ij}\mu_{ij}}\\
		\text{Average wait time:} \quad W_i &= \frac{\rho_{ij}}{c_{ij}\mu_{ij} - \lambda_i} + \frac{1}{\mu_{ij}} \quad (\text{when } \rho_{ij} < c_{ij})\\
		\text{Maximum throughput:} \quad \lambda_{max,i} &= \sum_j c_{ij}\mu_{ij}\\
		\text{Quality-adjusted throughput:} \quad \lambda_{quality,i} &= \lambda_{max,i} \times \text{AverageAccuracy}_i
	\end{align}
	
	\textbf{Enhanced Scalability Constraints}:
	\begin{itemize}
		\item Layer 1: Limited by computational resources ($\sim$2000/day)
		\item Layer 2: Limited by community size ($\sim$1200/day with 2000 active validators)
		\item Layer 3: Limited by expert availability ($\sim$150/day with 200 experts)
		\item Layer 4: Limited by cultural diversity ($\sim$80/day with global validator network)
		\item Layer 5: Limited by impact assessment complexity ($\sim$30/day with specialized assessors)
	\end{itemize}
	
	\subsubsection{Enhanced Validation Accuracy Modeling with Security Considerations}
	
	\textbf{Advanced Bayesian Validation Model with Security Weights}:
	\begin{align}
		\text{True contribution quality:} \quad Q &\sim \text{Beta}(\alpha, \beta)\\
		\text{Validator assessment:} \quad A_i | Q &\sim \mathcal{N}(Q, \sigma_i^2)\\
		\text{Security-adjusted assessment:} \quad A'_i &= A_i \times \text{SecurityWeight}_i
	\end{align}
	
	\textbf{Enhanced Posterior Quality Estimate}:
	\begin{equation}
		Q_{posterior} = \frac{\sum_i (A'_i/\sigma_i^2) \times \text{ReputationWeight}_i + \alpha}{\sum_i (1/\sigma_i^2) \times \text{ReputationWeight}_i + \alpha + \beta}
	\end{equation}
	
	\textbf{Multi-Dimensional Confidence Assessment}:
	\begin{align}
		\text{Technical Confidence:} \quad C_T &= 1 - \text{Var}(\text{TechnicalAssessments})\\
		\text{Social Confidence:} \quad C_S &= \text{CommunityConsensusLevel}\\
		\text{Security Confidence:} \quad C_{Sec} &= 1 - \text{Pr}(\text{Fraud Detected})\\
		\text{Overall Confidence:} \quad C &= \sqrt{C_T \times C_S \times C_{Sec}}
	\end{align}
	
	\subsection{Enhanced Economic Sustainability Analysis}
	
	\subsubsection{Comprehensive Cost-Benefit Framework with Validator Incentives}
	
	\textbf{Enhanced Validation Cost Per Contribution}:
	\begin{equation}
		\text{Cost} = \sum_{\text{layers}} (\text{ValidatorRewards} + \text{SecurityCosts} + \text{InfrastructureCosts} + \text{IncentiveCosts})
	\end{equation}
	
	\textbf{Updated Cost Estimates}:
	\begin{itemize}
		\item Layer 1: \$0.15 per contribution (algorithmic + security monitoring)
		\item Layer 2: \$8.00 per contribution (community validators + fraud detection)
		\item Layer 3: \$75.00 per contribution (expert validators + peer review)
		\item Layer 4: \$120.00 per contribution (cultural ambassadors + cross-cultural coordination)
		\item Layer 5: \$200.00 per contribution (impact assessors + longitudinal tracking)
	\end{itemize}
	
	\textbf{Enhanced Revenue Model with Multiple Streams}:
	\begin{align}
		\text{Revenue} &= \text{TokenRevenue} + \text{ServiceRevenue} + \text{PartnershipRevenue} + \text{DataRevenue}\\
		\text{TokenRevenue} &= \text{TokenPrice} \times \text{TokensIssued} \times \text{MarketDemand}\\
		\text{ServiceRevenue} &= \text{PremiumSubscriptions} + \text{EnterpriseServices}\\
		\text{PartnershipRevenue} &= \text{AcademicPartnerships} + \text{CorporateSponsorship}\\
		\text{DataRevenue} &= \text{AnonymizedDataSales} + \text{ResearchPartnerships}
	\end{align}
	
	\subsubsection{Validator Economics and Sustainability}
	
	\textbf{Validator Lifetime Value Model}:
	\begin{equation}
		\text{ValidatorLTV} = \sum_{t=0}^{T} \frac{\text{ValidatorValue}_t}{(1 + r)^t} - \sum_{t=0}^{T} \frac{\text{ValidatorCost}_t}{(1 + r)^t}
	\end{equation}
	
	Where:
	\begin{align}
		\text{ValidatorValue}_t &= \text{ContributionsValidated}_t \times \text{AverageContributionValue}\\
		\text{ValidatorCost}_t &= \text{Rewards}_t + \text{Support}_t + \text{Infrastructure}_t
	\end{align}
	
	\textbf{Network Effects and Enhanced Growth Model}:
	\begin{equation}
		\frac{dUsers}{dt} = \alpha \cdot Users \cdot (1 - \frac{Users}{CarryingCapacity}) \cdot NetworkEffect - \beta \cdot Users \cdot ChurnFactor
	\end{equation}
	
	Where:
	\begin{align}
		NetworkEffect &= 1 + \gamma \log(Users)\\
		ChurnFactor &= \begin{cases}
			\text{BaseChurn} & \text{if } TokenValue > MinViableValue \\
			\text{BaseChurn} \times 2 & \text{otherwise}
		\end{cases}
	\end{align}
	
	\section{Enhanced Implementation Architecture and Deployment Strategy}
	
	\subsection{Advanced System Architecture}
	
	\subsubsection{Enhanced Core Components with Security Integration}
	
	The proposed system architecture integrates multiple specialized components designed to work cohesively in a distributed validation environment. The architecture emphasizes security, scalability, and user experience while maintaining the integrity of the validation process. Each component serves a specific function within the larger ecosystem, and their interactions are carefully orchestrated to ensure optimal performance and reliability.
	
	The system's core foundation rests on five primary components that handle different aspects of the validation workflow. These components are supported by three specialized security modules that provide comprehensive protection against various threat vectors. Additionally, the architecture incorporates multiple storage solutions and user interfaces to accommodate diverse user needs and technical requirements.
	
	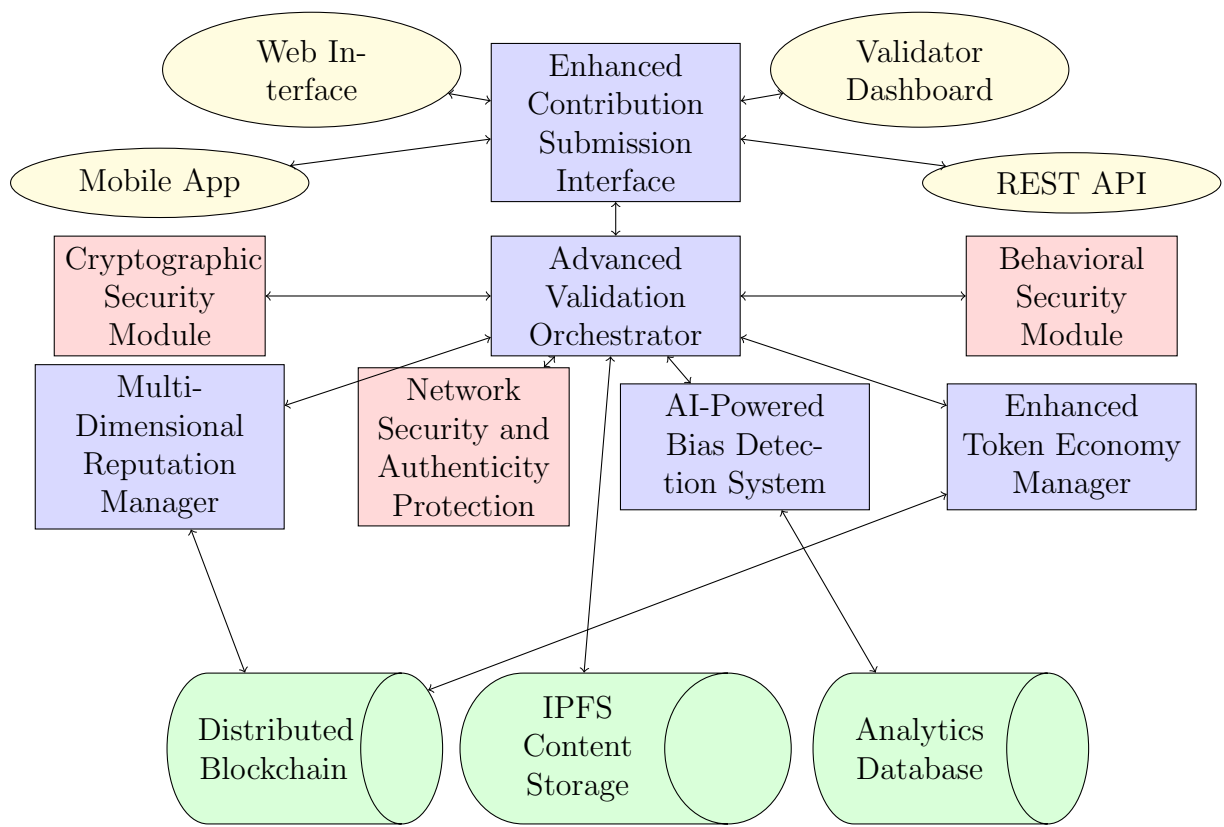
\begin{figure}[ht]
		\centering
		\begin{tikzpicture}[
			component/.style={rectangle, draw, fill=blue!15, minimum height=1.2cm, minimum width=3cm, text centered, text width=3cm},
			security/.style={rectangle, draw, fill=red!15, minimum height=1cm, minimum width=2.5cm, text centered, text width=2.5cm},
			storage/.style={cylinder, draw, fill=green!15, minimum height=1cm, minimum width=2cm, text centered, text width=2cm},
			interface/.style={ellipse, draw, fill=yellow!15, minimum height=0.8cm, minimum width=2.5cm, text centered, text width=2.5cm}
			]
			
			\node[component] (submission) at (0,4.3) {Enhanced Contribution Submission Interface};
			\node[component] (orchestrator) at (0,2) {Advanced Validation Orchestrator};
			\node[component] (reputation) at (-6,0) {Multi-Dimensional Reputation Manager};
			\node[component] (token) at (6,0) {Enhanced Token Economy Manager};
			\node[component] (bias) at (1.7,0) {AI-Powered Bias Detection System};
			
			\node[security] (crypto) at (-6,2) {Cryptographic Security Module};
			\node[security] (behavioral) at (6,2) {Behavioral Security Module};
			\node[security] (network) at (-2,0) {Network Security and Authenticity Protection};
			
			\node[storage] (blockchain) at (-4.5,-4) {Distributed Blockchain};
			\node[storage] (ipfs) at (-0.5,-4) {IPFS Content Storage};
			\node[storage] (analytics) at (4,-4) {Analytics Database};
			
			\node[interface] (web) at (-4,5) {Web Interface};
			\node[interface] (mobile) at (-6,3.5) {Mobile App};
			\node[interface] (api) at (6,3.5) {REST API};
			\node[interface] (validator) at (4,5) {Validator Dashboard};
			
			\draw[<->] (submission) -- (orchestrator);
			\draw[<->] (orchestrator) -- (reputation);
			\draw[<->] (orchestrator) -- (token);
			\draw[<->] (orchestrator) -- (bias);
			
			\draw[<->] (crypto) -- (orchestrator);
			\draw[<->] (behavioral) -- (orchestrator);
			\draw[<->] (network) -- (orchestrator);
			
			\draw[<->] (reputation) -- (blockchain);
			\draw[<->] (token) -- (blockchain);
			\draw[<->] (bias) -- (analytics);
			
			\draw[<->] (web) -- (submission);
			\draw[<->] (mobile) -- (submission);
			\draw[<->] (api) -- (submission);
			\draw[<->] (validator) -- (submission);
			
			\draw[<->] (orchestrator) -- (ipfs);
			
		\end{tikzpicture}
		\caption{Enhanced System Architecture with Integrated Security and Multi-Interface Support}
	\end{figure}
	
	The Enhanced Contribution Submission Interface represents the primary entry point for users interacting with the system. This component supports multiple content formats including code repositories, technical documents, and multimedia presentations. The interface provides real-time validation feedback to help contributors understand potential issues before final submission. To enhance user engagement, the submission process incorporates gamification elements with progress tracking and achievement unlocks. Integrated plagiarism detection and quality pre-checks ensure that only high-quality contributions enter the validation pipeline.
	
	The Advanced Validation Orchestrator serves as the central coordination hub for all validation activities. This component employs intelligent routing algorithms that analyze contribution type, complexity, and current system load to optimize validator assignment. Load balancing capabilities distribute validation tasks across available validator pools to maintain consistent response times. Real-time performance monitoring allows the system to identify bottlenecks and automatically adjust parameters for optimal throughput. When problematic cases arise that exceed normal validation parameters, automated escalation procedures ensure appropriate expert attention.
	
	The Multi-Dimensional Reputation Manager maintains comprehensive profiles for all system participants across multiple evaluation criteria. The system tracks behavioral patterns over time to identify trends in user and validator performance. When reputation scores decline due to temporary issues or misunderstandings, built-in recovery mechanisms provide pathways for users to rebuild their standing. Advanced algorithms optimize validator matching by considering domain expertise, cultural background, and historical performance metrics to maximize validation accuracy.
	
	The Enhanced Token Economy Manager oversees all economic aspects of the system through sophisticated reward calculation algorithms that account for contribution complexity, validation quality, and system needs. Dynamic inflation control mechanisms adjust token supply based on usage patterns and economic conditions to maintain stable value. Integration with achievement systems provides additional incentive structures beyond basic compensation. Cross-chain compatibility ensures that tokens can be utilized across multiple blockchain networks and platforms.
	
	The AI-Powered Bias Detection System continuously monitors validation patterns to identify potential bias in real-time. Cultural sensitivity analysis examines contributions and validations for content that might be inappropriate or exclusive in different cultural contexts. When bias is detected, the correction recommendation engine suggests specific improvements to address identified issues. A comprehensive fairness metrics dashboard provides transparency into system performance across different demographic and cultural groups.
	
	\subsubsection{Enhanced Scalability Considerations}
	
	The system's scalability strategy addresses both technical and organizational challenges inherent in distributed validation networks. Geographic distribution forms the foundation of the scaling approach, with regional validator specialization allowing for culturally appropriate and time-zone optimized validation services. This geographic distribution reduces latency and improves user experience while ensuring that cultural nuances are properly understood and addressed.
	
	The technical architecture employs a domain-specific microservices design that allows individual components to scale independently based on demand. Asynchronous processing with intelligent queue management ensures that system performance remains consistent even during peak usage periods. Auto-scaling mechanisms monitor contribution volume and complexity in real-time, automatically provisioning additional resources when needed. Cross-region validator collaboration protocols enable seamless cooperation between geographically distributed teams while maintaining security and data integrity.
	
	Performance targets for the enhanced system reflect ambitious but achievable goals based on current distributed system capabilities. The system aims to process between 1500 and 3000 contributions daily while maintaining average validation times between 0.5 and 2 days depending on priority level. System uptime targets of 99.9\% ensure reliable service availability, while validator accuracy targets above 85\% across all validation layers maintain quality standards. Security incident response time targets of less than one hour demonstrate the system's commitment to rapid threat mitigation. User satisfaction metrics target above 85\% retention rates to ensure long-term sustainability and growth.
	
	\subsection{Comprehensive Deployment Phases with Validator Onboarding}
	
	\subsubsection{Phase 1: Enhanced Proof of Concept (8-15 months)}
	
	The initial deployment phase focuses on establishing core system functionality within a controlled environment. Open source software contributions provide the target domain for this phase, as they offer well-defined quality criteria and an existing community of potential validators. The system will support between 200 and 800 users with 50 to 150 validators during this phase, providing sufficient scale to test core functionality while remaining manageable for detailed monitoring and optimization.
	
	Validator onboarding represents a critical success factor for the entire system. Comprehensive training modules address each validator type's specific requirements and responsibilities. The mentorship program pairs experienced validators with newcomers to provide personalized guidance and support. A dedicated practice validation environment allows new validators to gain experience with simulated contributions before handling real submissions. The achievement unlocking system guides validators through increasingly complex validation scenarios, ensuring steady skill development. Regular feedback sessions with validators provide valuable insights for system improvements and help maintain high engagement levels.
	
	Success metrics for Phase 1 emphasize stability and foundational quality rather than ambitious growth targets. System stability above 98\% uptime ensures reliable operation during the critical early period. Validation accuracy targets above 80\% across all layers establish baseline quality standards. User satisfaction metrics targeting above 75\% retention after three months indicate successful user experience design. Validator engagement metrics targeting above 60\% active monthly participation ensure adequate validation capacity. Security incident targets of zero major breaches and fewer than five minor issues per month establish security baseline expectations.
	
	\subsubsection{Phase 2: Multi-Domain Expansion with Advanced Gamification (1.5-3 years)}
	
	The second phase introduces significant complexity through multi-domain expansion. Data analysis, technical documentation, and educational content join software contributions as supported domains. This expansion tests the system's ability to maintain quality and fairness across diverse contribution types with different evaluation criteria. The user base expands to between 2000 and 8000 users, while the validator pool grows to between 500 and 1500 participants across all domains.
	
	Advanced validator features introduced during this phase significantly enhance the user experience and validator retention. Specialized validator tracks recognize domain expertise and provide clear advancement pathways. Advanced achievement systems incorporate rare collectible badges and exclusive recognition programs. Validator tournaments and competitions introduce substantial rewards and foster healthy competition. Cross-cultural exchange programs for cultural ambassadors promote global understanding and reduce bias. Professional development partnerships with universities and companies provide valuable career advancement opportunities for dedicated validators.
	
	Success metrics for Phase 2 reflect the increased complexity and scale of operations. Multi-domain validation accuracy averaging above 82\% demonstrates the system's ability to maintain quality across diverse content types. Cultural bias metrics within acceptable bounds, defined as less than 15\% deviation from baseline fairness measures, ensure equitable treatment across cultural groups. Economic sustainability indicators showing positive validator return on investment confirm the system's long-term viability. Validator advancement pathways with above 40\% progression to higher tiers indicate successful professional development programs. Community growth metrics targeting sustained 15\% monthly growth demonstrate healthy ecosystem expansion.
	
	\subsubsection{Phase 3: Global Scale-Up with Full Ecosystem (3-7 years)}
	
	The final phase represents full system maturation with comprehensive multi-domain operation and advanced gamification features. The system supports over 20,000 users with more than 5000 validators across all categories. At this scale, the focus shifts to long-term sustainability, global impact measurement, and continuous innovation to maintain competitive advantage.
	
	Full ecosystem features introduced in this phase leverage advanced technologies and established partnerships. Complete achievement and ranking systems span all domains and provide comprehensive recognition for contributions and validation excellence. Virtual reality validation environments enable immersive evaluation of complex contributions that benefit from spatial or interactive assessment. AI-assisted validation tools enhance efficiency while maintaining human oversight and decision-making authority. Integration with major educational and research institutions provides academic credibility and access to expert validators. Global validator conferences and recognition programs foster community building and professional development. Advanced research collaborations and academic partnerships enable the system to contribute to broader knowledge advancement while benefiting from cutting-edge research insights.
	
	\section{Comprehensive Risk Analysis and Enhanced Mitigation Strategies}
	
	Risk assessment for scalable distributed validation systems requires careful analysis of both technical and economic factors. The interdependence between system quality, validator satisfaction, and user adoption creates complex feedback loops that can either reinforce success or accelerate failure. Understanding potential failure modes and developing comprehensive recovery protocols ensures system resilience during crisis situations.
	
	\subsection{Primary Risk Factors with Quantified Impact}
	
	\subsubsection{Scalability and Validator Economics}
	
	Validator Quality Dilution represents the most significant scalability risk with high probability (0.75) and severe potential impact. As the system scales and demand for validators increases, the risk of accepting lower-quality validators to meet capacity needs becomes substantial. If system accuracy drops below 70\%, the resulting credibility crisis could lead to quantified losses between \$500,000 and \$2,000,000 in token value decline, accompanied by a 40-60\% user exodus.
	
	The mitigation approach for quality dilution must be comprehensive and proactive. Mandatory continuous education programs for all validators ensure that skills remain current as the system evolves, while regular competency testing with automatic tier adjustments maintains quality standards and provides clear performance feedback. Peer mentorship programs linking experienced validators with newcomers accelerate skill development and maintain institutional knowledge transfer. AI-assisted validation tools help catch quality degradation early before it impacts user experience, and economic incentives structured to reward long-term quality over short-term volume align validator interests with system success.
	
	Validator Economic Sustainability Crisis presents medium-high probability (0.65) with potentially catastrophic impact on system operation. If validators cannot earn adequate compensation for their time and effort, mass exodus could reduce system throughput to below 100 contributions per day, effectively rendering the platform unusable. This economic vulnerability requires multiple protective measures working in concert.
	
	Dynamic reward adjustment algorithms based on market conditions help maintain competitive compensation levels, while multiple revenue streams including external sponsorship and partnership opportunities reduce dependence on token economics alone. A validator insurance fund guaranteeing minimum compensation levels provides security during economic downturns, and performance-based bonuses that scale with system success align validator rewards with platform growth. Non-monetary benefits including professional development opportunities and industry recognition provide additional value beyond direct compensation, creating a more sustainable validator ecosystem.
	
	\subsubsection{Security and Fraud Threats}
	
	Security risks in distributed validation systems extend beyond traditional cybersecurity concerns to include sophisticated social engineering and coordinated manipulation attempts. Sophisticated Fraud Ecosystem Development carries medium probability (0.55) but could compromise fundamental system integrity. Organized fraud rings might develop systematic approaches to game validation processes, potentially making 20-40\% of contributions fraudulent without detection.
	
	The defense against sophisticated fraud requires advanced technological solutions combined with proactive detection mechanisms. Advanced machine learning models trained on fraud patterns help identify suspicious behavior before it scales, while honeypot contributions designed to attract fraudulent activity help identify bad actors early in their operations. Cross-validation with external reputation systems provides independent verification of user credibility, and real-time anomaly detection with automatic flagging prevents fraud from spreading through the system. A comprehensive legal framework for pursuing large-scale fraud operations provides deterrent effects and recovery mechanisms for cases that bypass technological defenses.
	
	Cultural and Social Engineering Attacks represent medium probability (0.50) threats that could systematically bias the platform toward particular viewpoints or exclude minority perspectives. Such attacks could reduce global adoption and severely damage the platform's credibility in diverse markets. The impact extends beyond immediate user loss to long-term reputation damage in international markets where cultural sensitivity is paramount.
	
	Defending against cultural manipulation requires both structural safeguards and ongoing monitoring. Mandatory cultural diversity requirements for validation pools ensure balanced perspective representation, while anonymous validation options for sensitive contributions protect against cultural bias in controversial topics. Regular bias audits by external cultural competency experts provide independent assessment of system fairness, and validator rotation policies prevent long-term influence accumulation by any single group. Transparent bias metrics published in real-time dashboards maintain public accountability for system fairness and allow the community to monitor potential manipulation attempts.
	
	\subsection{Enhanced Failure Mode Analysis with Recovery Protocols}
	
	The most critical failure modes represent combinations of individual risks that create cascading effects throughout the system. Each failure mode requires specific recovery protocols that address both immediate symptoms and underlying causes.
	
	Cascade Quality Degradation represents a critical failure mode triggered when multiple high-reputation validators simultaneously reduce their effort levels. This creates a downward spiral where lower accuracy leads to reduced token value, causing more validators to leave, which further reduces accuracy in a self-reinforcing cycle. The cascade effect can transform a minor quality issue into system-wide credibility collapse within weeks if not addressed immediately.
	
	The enhanced recovery protocol for quality degradation begins with immediate activation of validator retention incentives, doubling rewards for 30 days to maintain critical validator participation during the crisis period. Emergency recruitment of expert validators from partner institutions provides temporary capacity while the system stabilizes, and temporary increases in algorithmic pre-screening reduce human validation load during the recovery phase. System parameter adjustments requiring fewer validators per contribution maintain throughput with reduced capacity, while a comprehensive community communication campaign explains recovery measures and provides realistic timelines for system restoration.
	
	Cultural Fragmentation with Fraud Exploitation occurs when cultural values conflicts escalate while fraud rings exploit the resulting divisions. This represents a particularly dangerous combination where the system splits into cultural silos, reducing network effects while fraud exploitation increases in the confused environment. The dual nature of this crisis requires coordinated responses addressing both cultural harmony and security simultaneously.
	
	Recovery protocols for cultural fragmentation include activation of emergency cross-cultural mediation protocols facilitated by trained cultural ambassadors who understand both the technical system and cultural dynamics. Temporary implementation of cultural quota systems for validation ensures balanced representation during the crisis period, while enhanced fraud detection specifically targeting exploitation of cultural divisions prevents bad actors from taking advantage of system instability. Community dialogue sessions facilitated by cultural ambassadors help resolve underlying conflicts, and governance reforms better accommodate cultural diversity while maintaining system integrity and preventing future fragmentation.
	
	Security Breach with Validator Compromise represents the most severe potential failure mode, where a major security incident compromises validator accounts and validation integrity simultaneously. This triggers immediate trust collapse, user exodus, and potential complete system credibility destruction. The compound nature of this crisis, affecting both security and validation quality, requires the most comprehensive recovery approach.
	
	The recovery protocol begins with immediate system halt and comprehensive security audit activation to contain the breach and prevent further damage. All validator accounts undergo verification procedures with potential reset of compromised credentials, while enhanced multi-factor authentication implementation for all validators prevents similar future breaches. Third-party security audits and penetration testing identify and address all vulnerabilities discovered during the incident investigation. Gradual system restart with enhanced monitoring and reduced complexity ensures stability during recovery, and a comprehensive transparency report combined with a compensation fund for affected users demonstrates accountability and commitment to user protection.
	
	The interconnected nature of these risks requires continuous monitoring and adaptive response capabilities. Success depends not only on preventing individual risks but also on recognizing and responding to the complex interactions between different risk factors. The recovery protocols serve as both reactive measures for crisis response and proactive frameworks for building system resilience against future challenges.
	
	\section{Conclusion}
	
	This paper presents EarthOL, a comprehensive theoretical and practical framework for replacing computational waste in blockchain systems with meaningful human contributions within carefully bounded domains. Our analysis reveals that while universal value assessment remains impossible, significant progress is achievable in high-consensus domains through domain restriction, cultural bias mitigation, advanced security protocols, and sophisticated validator incentive systems.
	
	\subsection{Comparative Analysis with Existing Systems}
	
	To understand EarthOL's position in the current landscape of consensus mechanisms, we provide a comprehensive comparison with existing systems across multiple dimensions including energy consumption, social impact, throughput, decentralization, fairness, and gamification capabilities.
	
	\begin{table}[ht]
		\centering
		\resizebox{\textwidth}{!}{ 
			\small
			\begin{tabular}{|l|c|c|c|c|c|c|}
				\hline
				\textbf{System} & \textbf{Energy} & \textbf{Social} & \textbf{Throughput} & \textbf{Decentralization} & \textbf{Fairness} & \textbf{Gamification} \\
				\hline
				Bitcoin PoW & Very High & None & Low & Very High & Medium & None \\
				Ethereum PoS & Low & None & Medium & High & Low & None \\
				Filecoin & Medium & Low & Medium & High & Medium & None \\
				Steemit & Low & Medium & Low & Medium & Low & Basic \\
				\textbf{EarthOL PoHC} & \textbf{Low} & \textbf{Very High} & \textbf{Low-Medium} & \textbf{Medium-High} & \textbf{Medium-High} & \textbf{Advanced} \\
				\hline
		\end{tabular}}
		\caption{Enhanced Comparison of Consensus Mechanisms with Gamification Features}
	\end{table}
	
	This comparison demonstrates that EarthOL occupies a unique position in the consensus mechanism landscape, prioritizing social impact and human engagement over raw computational throughput. While traditional proof-of-work systems like Bitcoin achieve maximum decentralization at the cost of enormous energy consumption, and proof-of-stake systems like Ethereum improve energy efficiency but provide limited social benefits, EarthOL's Proof of Human Contribution represents a paradigm shift toward meaningful human participation in blockchain consensus.
	
	\subsection{Theoretical Foundations and Social Choice Theory}
	
	Our approach extends beyond traditional mechanism design by addressing fundamental challenges in social choice theory while acknowledging the boundaries established by Arrow's impossibility theorem. While Arrow's theorem applies to general preference aggregation, our domain-restricted approach with gamified incentives opens new possibilities for meaningful progress in specific contexts. The system provides intrinsic motivation through achievement systems that complement economic incentives, creates social learning environments where validators improve judgment through peer interaction, implements progressive complexity systems that gradually prepare validators for difficult decisions, and establishes cultural bridge-building through gamified cross-cultural collaboration.
	
	The framework extends traditional mechanism design by incorporating social gaming elements that provide non-monetary motivation, enabling long-term reputation building that creates sustained engagement. Community-driven governance adapts to changing social preferences, while multi-dimensional reward systems recognize diverse forms of contribution. This comprehensive approach addresses both the theoretical limitations of preference aggregation and the practical challenges of maintaining human engagement in decentralized systems.
	
	\subsection{System Limitations and Boundaries}
	
	Despite its advantages, EarthOL faces several fundamental limitations that must be acknowledged. From a theoretical perspective, the system cannot overcome Arrow's impossibility theorem in the general case, though gamification may increase the domain of feasible preference aggregations. Cultural bias reduction has theoretical lower bounds, but social learning mechanisms may gradually shift these bounds over time. Scalability remains fundamentally constrained by human cognitive capacity, though AI augmentation may extend practical limits. Additionally, gaming incentives may create new forms of strategic behavior not fully captured in current theoretical models.
	
	Practical constraints present additional challenges that require careful management. The system requires sustained high-quality validator participation, which depends on long-term economic and social incentive alignment that may be difficult to maintain. The economic model's sustainability depends on external token value and broader ecosystem health, creating dependencies on factors beyond the system's direct control. Cultural coordination costs grow exponentially with diversity, potentially limiting global scalability in highly heterogeneous environments. Furthermore, gamification elements require constant innovation to maintain engagement over time, necessitating ongoing development resources.
	
	\subsection{Key Theoretical Contributions}
	
	This research makes several significant theoretical contributions to the field of decentralized consensus mechanisms. We present a mathematical framework for domain feasibility assessment that incorporates gamification considerations, enabling systematic evaluation of which domains are suitable for human-centered validation. Our game-theoretic analysis of human-centered consensus stability examines multi-dimensional incentives and their effects on system equilibrium. The probabilistic modeling of cross-cultural validation accuracy includes security adjustments that account for various attack vectors and cultural biases.
	
	The economic sustainability analysis for human-verification systems provides a comprehensive framework for understanding validator economics and long-term system viability. Our advanced security framework addresses sophisticated fraud and cultural attack vectors that are unique to human-centered systems. Additionally, we develop comprehensive validator motivation and engagement models that incorporate both social and economic factors, providing insights into maintaining human participation in decentralized systems over extended periods.
	
	\subsection{Practical Implications and Implementation Insights}
	
	The practical implications of this research are substantial and actionable. Domains with high consensus levels (feasibility factor $F > 0.7$) show strong promise for implementation when combined with proper validator incentives and engagement mechanisms. The five-layer verification architecture with integrated security measures successfully balances efficiency, accuracy, and safety requirements while maintaining the human-centered approach that distinguishes EarthOL from purely computational systems.
	
	A gradual deployment strategy with comprehensive validator onboarding minimizes implementation risks while allowing for iterative improvement based on real-world feedback. Economic sustainability appears achievable through diversified revenue streams and careful parameter management, though this requires ongoing monitoring and adjustment. Advanced gamification systems can maintain long-term validator engagement while improving overall system quality, creating a positive feedback loop between participation and system effectiveness.
	
	\subsection{System Capabilities and Performance Boundaries}
	
	EarthOL's enhanced capabilities demonstrate significant improvements over earlier human-centered consensus approaches. Throughput scalability reaches approximately $3 \times 10^3$ contributions per day with enhanced validator pools and AI assistance, representing a substantial improvement in processing capacity while maintaining human judgment quality. Cultural bias can be reduced to acceptable levels with deviation rates below 15\% through advanced mitigation strategies that combine algorithmic approaches with social learning mechanisms.
	
	The system's applicability extends to an expanding set of domains as cultural consensus grows and technology improves, suggesting that the framework's utility will increase over time. Security can be maintained against sophisticated attacks through multi-layer defense systems that address both technical vulnerabilities and social manipulation attempts. Validator satisfaction and retention appear achievable through comprehensive incentive and recognition systems that address both economic and social motivations for participation.
	
	\subsection{Future Research Directions}
	
	The research agenda emerging from this work spans multiple time horizons and research domains. Near-term research priorities include large-scale prototype development with over 1000 validators across multiple domains to validate theoretical predictions in real-world conditions. Comprehensive cross-cultural value assessment experiments in controlled environments will help refine our understanding of cultural bias mitigation strategies. Economic sustainability modeling using real market data and validator feedback will inform practical deployment decisions, while advanced fraud detection algorithm development using machine learning will strengthen security frameworks. Longitudinal studies of validator motivation and engagement in gamified environments will provide crucial insights for maintaining system effectiveness over time.
	
	Medium-term research directions focus on system integration and optimization. AI-human hybrid validation systems that maintain human agency while improving efficiency represent a promising avenue for scaling human-centered consensus. Governance mechanism optimization through participatory design with validator communities will ensure that system evolution reflects user needs and preferences. Integration with existing social platforms and academic institutions could provide natural pathways for system adoption, while development of domain-specific validation protocols for emerging fields will expand the system's applicability. Cross-chain interoperability for validator reputation and contribution tracking will enable broader ecosystem integration.
	
	Long-term research opportunities address fundamental questions about human coordination and social systems. Large-scale social experiments with over 100,000 participants across multiple cultures will provide unprecedented insights into human consensus formation at scale. Investigation of cultural evolution effects on value consensus over decades will inform our understanding of how human coordination systems evolve over time. Development of global coordination mechanisms for human-centered systems addresses one of the most challenging problems in distributed systems design. Integration with emerging technologies like brain-computer interfaces for enhanced validation represents a frontier research area that could fundamentally transform human-computer interaction in consensus systems. Long-term impact studies on social behavior and cooperation at scale will help us understand the broader societal implications of human-centered blockchain systems.
	
	\subsection{Final Assessment and Future Outlook}
	
	EarthOL represents a significant step toward more socially beneficial consensus mechanisms that harness human creativity and judgment while maintaining the security and decentralization benefits of blockchain systems. The integration of gamification elements with serious validation work represents a novel approach to maintaining human engagement in decentralized systems while creating genuine social value. While the system cannot solve the general value assessment problem identified by Arrow's impossibility theorem, it provides a comprehensive framework for meaningful human-centered consensus in specific domains where cultural agreement is sufficient.
	
	The enhanced validator incentive systems and comprehensive security frameworks address many practical concerns about real-world implementation while opening new research directions in human-centered decentralized systems. The primary contribution lies in identifying and expanding the boundaries of what is possible in decentralized human value assessment while providing a mathematically grounded, security-conscious, and socially engaging framework for progress within those boundaries.
	
	Future validation of these theoretical predictions through careful empirical research, large-scale prototype development, and long-term studies of validator behavior and system evolution will be crucial for realizing the full potential of human-centered consensus mechanisms. The framework presented here provides a foundation for this future work while demonstrating that meaningful alternatives to energy-intensive computational consensus are both theoretically sound and practically achievable within appropriate domain boundaries.
	
	\section*{Acknowledgments}
	
	We acknowledge the fundamental challenges in human value assessment and recognize that our enhanced mathematical models, while more comprehensive, still represent simplifications of complex social phenomena. The theoretical frameworks presented require extensive empirical validation, large-scale testing, and long-term monitoring to be meaningfully implemented. We thank the global validator community for their anticipated contributions to system development and testing.
	
	\bibliographystyle{plain}

\end{document}